\documentclass[aps,pra,twocolumn,showpacs,amsmath,amssymb]{revtex4}

\usepackage{amsmath}
\usepackage{amssymb}
\usepackage[letterpaper,textwidth=6.0in,textheight=8.6in,left=1.55in,top=1.1in]{geometry}
\usepackage{graphicx}
\usepackage{epsf}
\usepackage{flafter}%
\usepackage{bm}

\begin{document}

\title{Many-body Rabi oscillations of Rydberg excitation in small mesoscopic samples}

\author{J. Stanojevic}
\author{R. C\^ot\'e}
\affiliation{Department of Physics, University of Connecticut,
Storrs, CT 06269}
\date{\today}

\begin{abstract}
We investigate the collective aspects of Rydberg excitation in ultracold mesoscopic systems. Strong interactions between Rydberg atoms influence the excitation process and impose correlations between excited atoms. The manifestations of the collective behavior of Rydberg excitation are the many-body Rabi oscillations, spatial correlations between atoms as well as the fluctuations of the number of excited atoms. We study these phenomena in detail by numerically solving the many-body Schr\"edinger equation. 
\end{abstract}


\maketitle

\section*{Introduction}
Long radiative lifetimes of Rydberg atoms and their possibility to interact strongly at large distances have made ultracold Rydberg atoms interesting  systems for possible quantum information applications. The electron in a Rydberg state is very far from the nucleus and thus sensitive to external fields or the presence of neighboring Rydberg atoms. Due to the huge polarizabilities of Rydberg atoms, it is possible to induce relatively large electric dipole moments using small electric fields. The capability to turn on and off the interactions just by switching the external field is an important aspect of this approach to quantum computation. In this way decoherence effects due to the interactions between atoms or with the environment, can be significantly reduced. Strong interactions can be used to entangle neutral atoms and achieve fast quantum gates \cite{jaksch00,Grangier02}, as well as to blockade excitation by shifting many-atom excited states out of resonance. It has been proposed to use this blockading effect to realize scalable quantum gates \cite{lukin01}. The evidence of excitation blockade has been found in several experiments with a narrow laser bandwidth. In \cite{tong04}, a local blockade of Rydberg state excitation in a mesoscopic sample due to strong van der Waals (vdW) interactions has been observed using a pulse-amplified single-mode laser. The $5s$ ground-state rubidium atoms were excited by one photon UV transitions to  high $np_{3/2}$ Rydberg states. Rydberg excitation exhibited dramatic suppression compared to the non-interacting case. A mean-field type model was proposed to explain these experimental results. In the model, different atoms interact differently depending on their locations. Different interaction energies were modeled by a distribution of mean-field shifts for which a distribution of excitation probabilities was calculated. A good agreement between the theoretical model and experimental measurements was found.Significant suppression of Rydberg excitation has also been observed using cw excitation \cite{singer04}. This Rydberg excitation, strongly influenced by interactions, exhibits sub-Poissonian atom counting statistics \cite{Liebisch,Ates06}. The blockade effect due to dipole-dipole interactions in an ultracold sample of Cs atoms \cite{voght06} has been reported. Also, an interesting antiblockade effect in two-step excitation processes was predicted \cite{Ates07,Pohl07}.

At high principal quantum number $n$, the interactions between Rydberg atoms are quite strong and they inhibit the excitation of many surrounding atoms within a range of few $\mu{\rm m}$.  The atoms within this range are strongly correlated so that a many-body treatment is, in general, needed. In \cite{Hernandez} the many-body wave function was calculated by numerically solving the Schr\" odinger equation.  This type of analysis can be important for quantum information applications \cite{Hernandez06} because precise control at the quantum level is essential in these applications. Strong interactions can affect the coherent manipulation of a large group of atoms so it is necessary to use a many-body treatment to evaluate the fidelity of quantum information protocols. 

In this paper we consider mesoscopic samples of $\sim$10 $\mu{\rm m}$ diameter containing up to 100 ground-state atoms which are excited by single photon transitions to high $np$ Rydberg states. 
We study in detail the dynamics of such systems, especially the possibility of many-body Rabi oscillations of Rydberg excitation.  It is plausible to investigate these oscillations in smaller systems because one would not normally expect to achieve the coherent manipulation of large groups of atoms. The many-body approach developed in \cite{Hernandez} is quite suitable for this analysis and we use it here. We only modify some technical details on how to treat interactions in this approach.

\section*{Many-body effect in ultracold Rydberg systems and collective oscillations}
Although models may be very useful in explaining some important overall properties of large strongly-interacting Rydberg systems, it is essential to include many-body correlations in the study of many-body effects. In ultracold Rydberg systems, the thermal motion of the atoms is greatly reduced so that in many situations it can be completely ignored. 

The idea of these collective oscillations can be explained as follows. If the interactions between atoms are strong enough, all many-body states with two or more excited atoms will be greatly shifted by these interactions and thus far-off resonance. Such systems are effectively two-level systems because there are only two (collective) states that are populated: the ground state $\left| G \right\rangle=\left| g_1g_2\ldots g_N \right\rangle$, where all atoms are in the atomic ground state $\left| g_i\right\rangle$, and only one excited state $\left| E \right\rangle=1/\sqrt{N}\sum_{i}\left| g_1\ldots e_i\ldots g_N \right\rangle$ where any, but only one,  atom can be excited. On atomic resonance such two-level systems are exactly solvable \cite{wolker}  and the solution, in terms of excitation probability $P_{\mathrm {exc}}$, is
\begin{equation}\label{full_block}
P_{\mathrm {exc}}(t)= \frac{1}{N}{\mathrm{sin}}^{2} \left(\sqrt{N}\Omega F(t)\right),
\end{equation}
where $\Omega F(t)$ is the pulse area and $N$ is the number of atoms in the sample. Clearly, the collective oscillations are much faster then the isolated-atom Rabi frequency $\Omega$. The question is whether these fast oscillations can exist in systems which are big enough that there can be several excited atoms, or alternatively with the interactions  not strong enough to fully blockade the system.

Because the atoms are ultracold, we ignore the effects of thermal motion during Rydberg excitation in this analysis. We consider the following many-body Hamiltonian of two-level atoms and Rydberg-Rydberg interactions ($\hbar=1$) 
\begin{eqnarray}\label{Hamiltonian}
H \!&=&\! \Delta \sum\limits_{i = 1}^N {\hat \sigma _{ee}^i }  + \frac{\Omega}{2}  \sum\limits_{i = 1}^N \left( {f(t) \hat \sigma _{eg}^i  +f^{*}(t) \hat \sigma _{ge}^i } \right)\nonumber\\
&&{} + \sum\limits_{i = 1,j > i}^N {\kappa_{ij}} \hat \sigma _{ee}^i \hat \sigma _{ee}^j \, ,\label{Hamiltonian}
\end{eqnarray}
where $\Delta$ is the frequency detuning and the interactions between Rydberg atoms are given by $\kappa_{ij}$. The second term in the Hamiltonian is the dipole operator representing the interaction with the optical field. The function $f(t)$ is the time evolution (envelope) of the laser pulse.  The non-trivial parts of the $\sigma$-operators  are defined as 
$\hat \sigma _{\alpha\beta}^i = \left| {\alpha_i } \right\rangle \left\langle {\beta_i } \right|$, where $\alpha,\beta$ reffer either to the ground state $g$ or the excited state $e$.

We can always eliminate the detuning from the previous Hamiltonian by the unitary transformation
$\exp\left(it\, \Delta \Sigma_{i = 1}^N {\hat \sigma _{ee}^i }\right)H\exp\left(-it\, \Delta \Sigma_{i = 1}^N {\hat \sigma _{ee}^i }\right)$. In the remaining part of the Hamiltonian, the only effect of this transformation is to replace $f(t)$ by $f(t) \exp(it\Delta)$. This transformation can simplify the use of other approximations applied to solve (\ref{Hamiltonian}). We assume that it is always performed and there is no need to consider this term explicitly. In fact, we set $\Delta=0$ in our simulation.

Since the dimensionality of the problem is $2^N$, which is a huge number for numerical calculations with $N\approx 100$, it is necessary to make some approximations. The idea of local blockade is utilized here (more explanations can be found in \cite{Hernandez}). We can group atoms in such a way that the probability to have two or more excited atoms within a group is practically zero. Such groups of atoms are often called superatoms. Each superatom ``i" is a two-level system described by two collective states  $\left| G_i \right\rangle$ and $\left| E_i \right\rangle$. 
In the next step, the actual many-body Hamiltonian is represented/approximated in terms of these superatoms. For convenience we define new operators $\sigma _{EG}^i=\left| E_i \right\rangle\left< G_i \right|$ and $\sigma _{EE}^i=\left| E_i \right\rangle\left< E_i \right|$. In terms of the new operators, the interaction with the laser field 
can be expressed as follows
\begin{eqnarray}
 &\sum\limits_{i = 1}^N \frac{\Omega}{2}&\left( {f(t) \hat \sigma _{eg}^i  +f^{*}(t) \hat \sigma _{ge}^i } \right)\label{laserField}\nonumber\\ 
&&\hspace{-9 mm}\rightarrow\sum\limits_{j = 1}^{N_{\bf sa}} \frac{\Omega\sqrt{N_i}}{2} \left( {f(t) \hat \sigma _{EG}^j  +f^{*}(t) \hat \sigma _{GE}^j } \right) , 
\end{eqnarray}
because
\begin{equation*}
\left< E_i  \right|\sum\limits_{j = 1}^N \!\left( {f(t) \hat \sigma _{eg}^j \! +\!f^{*}(t) \hat \sigma _{ge}^j } \right) \left| G_i  \right>\!=\! \frac{\sqrt{N_i}\Omega}{2}f(t),
\end{equation*}
where $N_{\bf sa}$ is the number of superatoms and $N_{j}$ is the number of atoms in a superatom. The Rydberg interaction $V_{Ryd}$ between superatoms ``i"  and ``j" in their excited states is
\begin{equation}\label{SDinteract}
\begin{split}
k_{ij}&=\left< E_i  E_j \right| V_{Ryd} \left| E_i  E_j \right>\\
&= \frac{1}{N_i N_j} \sum_{p\leq N_i}\sum_{q\leq N_j}  \kappa_{pq}.
\end{split}
\end{equation}
In these sums, atoms ``p"  and ``q" belong to different superatoms. The last formula gives the prescription for how to introduce the interaction between superatoms. It seems that the  interaction between two superatoms in \cite{Hernandez} was modeled using the distance between their centers of mass. Equation (\ref{SDinteract}) suggests that the superatom interaction is rather defined by  $\left<\kappa_{pq}\right>$ instead of $\kappa(\left<r_{pq}\right>)$.  Using $\kappa(\left<r_{pq}\right>)$ tends to underestimate the influence of interactions. The difference between these two ways can be significant if $\kappa(r)$ depends strongly on $r$, which is the case here. However, the effect of such differences in interaction energies on  the results of simulations is hard to judge because the largest $\kappa_{pq}$ do not contribute to the interactions between superatoms (explained later in the text)  and because of the strong suppression of Rydberg excitation. This $k_{ij}$, as the relevant parameter to describe the interactions between superatoms, should be preferably used in the process of the superatom formation as well. These are the only modifications we make to the method  \cite{Hernandez}. Another useful simplification comes from the suppression of Rydberg excitation. Namely, the number of excited atoms is limited due to the blockade effect and there is no need to consider many-body states with high numbers of excited atoms. To illustrate the size of the problem after these steps, we note that for systems with twenty-three superatoms and at most seven of them excited, there are about four hundred thousand excitation amplitudes to solve.
\begin{figure}
    \centerline{\epsfxsize=2.9in\epsfclipon\epsfbox{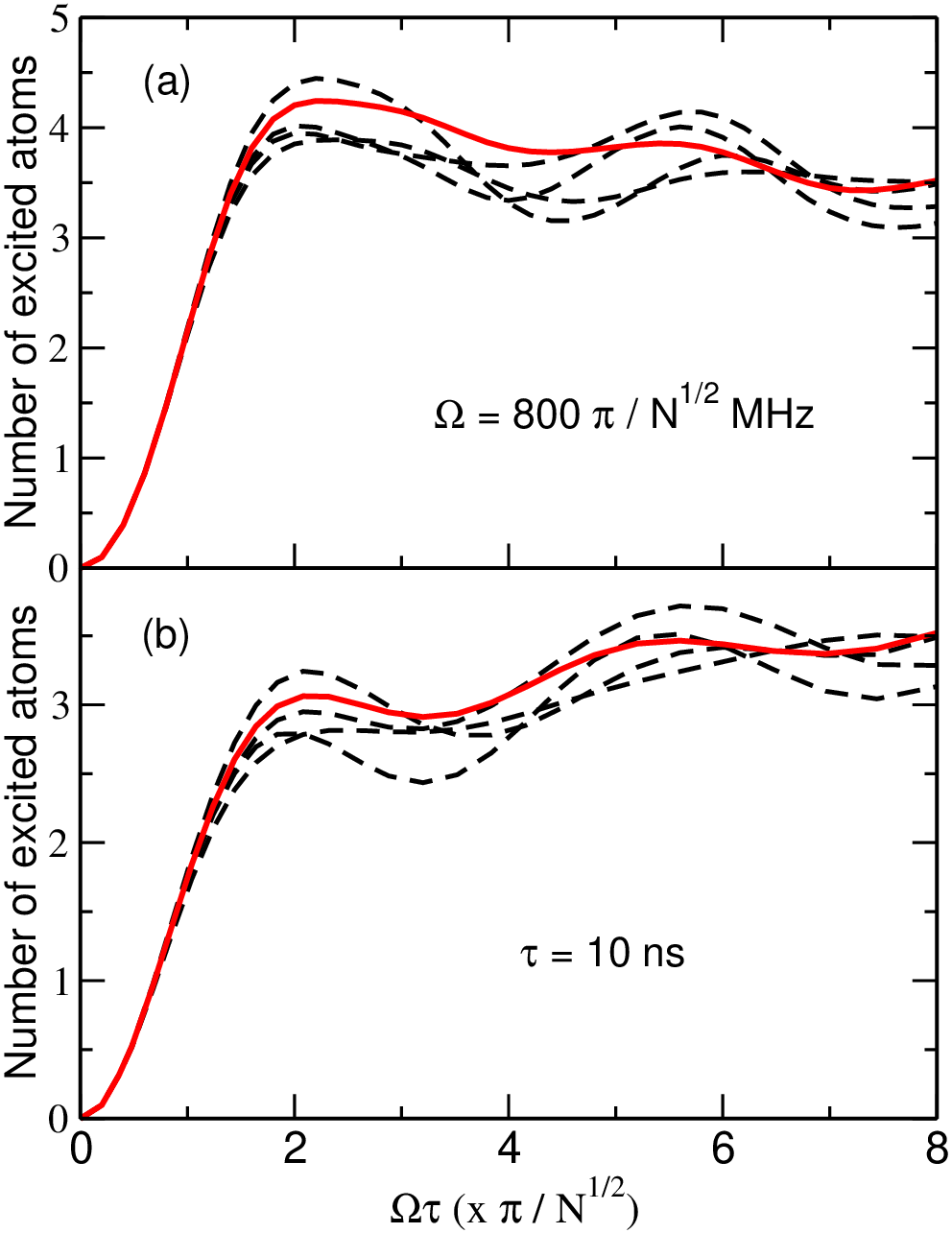}}
\caption{\label{fig_co1} The number of excited 70 $p_{3/2}$ atoms as a function of the pulse area for square laser pulses ($N=70$ and $\rho=10^{11}$ cm$^{-3}$).  In (a) the pulse duration was varied and in (b) the single-atom Rabi frequency. In both graphs the dashed lines represent some particular random distributions of atoms, while the solid lines are the average dependencies over 100 random arrangements of atoms.}
\end{figure}

To form superatoms we use the same recursion described in \cite{Hernandez}, except that the largest $k_{ij}$ is our guidance in deciding which (super)atoms to group together. We use the following recursion.  We initially set the number of superatoms to be equal to the number of atoms. After that, we start the recursion. In the first step, we calculate the interactions between superatoms using Eq. (\ref{SDinteract}). In the second step, we check if the number of superatoms is equal to the desired number (chosen in advance) of superatoms. If it is, the recursion is over, otherwise, we execute the third step. In this step, we group superatoms corresponding to the largest $|k_{ij}|$. Therefore, the number of superatoms is reduced by one and thus we need to recalculate all $k_{ij}$. This means that the recursion cycle is initiated again.

\section*{Results and discussion}
We consider mesoscopic samples of $\sim$10  $\mu{\rm m}$ diameter containing  up to 100 ground-state atoms. It is assumed that $5s$ ground state rubidium atoms are excited by one-photon transitions to $70p_{3/2}$ Rydberg states. The positions of atoms within a sample are randomly generated before the time evolution of the system takes place. We typically consider samples of 70 atoms and density of $10^{11}$ cm${}^{-3}$. The only exception is the analysis of density fluctuations, shown in Fig \ref{fig_co5}, where the number of atoms varies according to a Poissonian distribution. The effect of changing the density is similar to changing the interaction strength, and so we can vary only one of them. We vary the scaled interaction strength which is obtained as follows. From the Schr\"odinger equation $i\partial \psi/\partial t=H \psi$, after the scaling $t\rightarrow  t/\tau$, $\Omega\rightarrow  \Omega\tau$ and $k_{ij}\rightarrow  k_{ij} \tau$, we conclude that the final excitation probability $P_{\rm exc}$ is a function of the pulse area and the product $k_{ij} \tau\sim k_{ij}/\Gamma$, i.e. $P_{\rm exc}=P_{\rm exc} (\Omega \tau, k_{ij} \tau)$. Here $\tau$ is the pulse duration (for a Gaussian it is the FWHM) and $\Gamma$ is the bandwidth. In our simulations, the range of possible pulse areas ($\sim\Omega \tau$) is the same for all types of laser pulses. Because the single-atom Rabi frequency $\Omega$ is scaled by $\sqrt{N_i}$, in the regime where the correlations between atoms are significant, the many-body characteristics of Rydberg excitation should be present at relatively small $\Omega$. 

The effect of interactions can be amplified, not only by increasing $k_{ij}$, but also $\tau$. Increasing $\tau$ also means reducing the bandwidth (for ideal pulses). When, for some $\tau$, the interactions between nearest neighbors  become comparable with the bandwidth, the influence of interactions on the excitation dynamics should become noticeable. In our simulations the interactions between atoms are given by $k_{ij}=\tilde C_6 /R_{ij}^6$. This type of interactions is also supported by the experiment \cite{Amthor}, in which the production of ions in collision processes is consistent with the assumption of an attractive van der Waals potential and a theoretical estimate of its magnitude. In the model presented in \cite{tong04}, we calculated the effective vdW coefficient $\tilde C_6$ to be $7 C_6/60$, where $C_6$ is the dispersion coefficient of the strongest $np\!+\!np$ potential \cite{singer-jpb}. This $\tilde C_6$ is essentially model-dependent, however, using this value we had a good agreement with the experimental data \cite{tong04} and thus we use it in this calculation as well. 

In Fig. \ref{fig_co1}, the time and omega dependencies of the excitation probability are shown for square laser pulses in  panels (a) and (b), respectively. These two dependencies are presented as functions of the pulse area. In both graphs the dashed curves correspond to particular random distributions of atoms within the sample. The solid curves are the average over the results obtained from one hundred random arrangement of atoms. If we use fifty different placements instead of one hundred, the difference in the average excitation probability is only about 1 $\%$.  The evaluation of the omega dependence is much more demanding because for each value of $\Omega$ we have to calculate the corresponding time dependence first. Although excitation probabilities for particular random distributions of atoms clearly exhibit many-body oscillations, when averaged over many distributions, these oscillations are significantly suppressed. Collective oscillations presented in Fig. \ref{fig_co5}  are more pronounced because a much greater scaled interaction strength is used.

\begin{figure}
    \centerline{\epsfxsize=2.9 in\epsfclipon\epsfbox{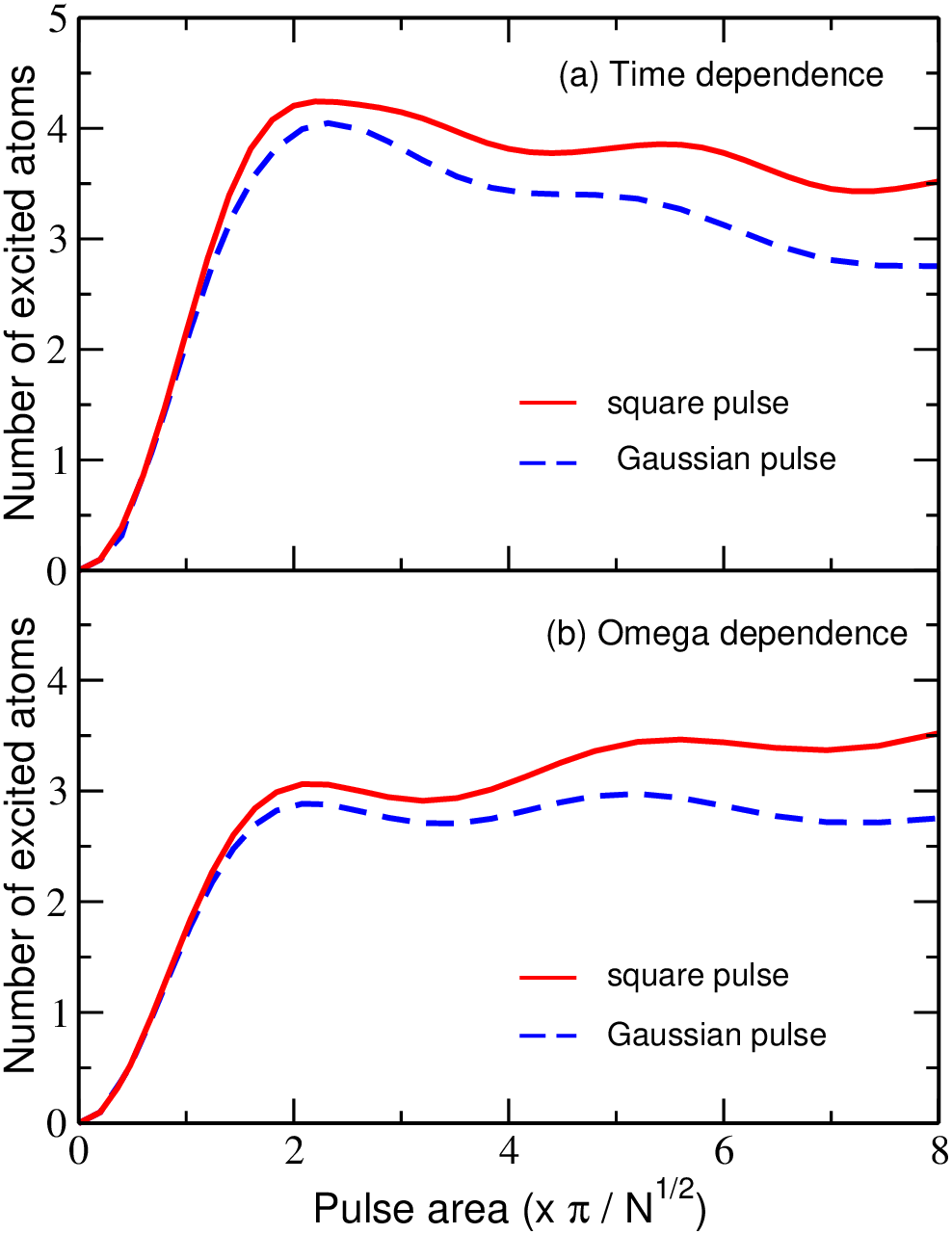}}
\caption{\label{fig_co2} The average dependencies of the number of excited 70 $p_{3/2}$ atoms versus the pulse area for different types of laser pulses ($N=70$ and $\rho=10^{11}$ cm$^{-3}$). In the absence of interactions all these dependencies should be the same. In (a) the pulse duration was varied and in (b) the single-atom Rabi frequency. Both solid lines correspond to square pulses and dashed lines to Gaussian ones. This plot demonstrates that the excitation blockade is a bit more efficient for Gaussian pulses.}
\end{figure}

In Fig. \ref{fig_co2} the average time and omega dependencies of the excitation probability for  square and Gaussian pulses are presented. Here we show how different pulse shapes affect Rydberg excitation. It is known that for resonant single-atom excitation, the excitation probabilities depend only on the pulse area. However, from $P_{\rm exc}=P_{\rm exc}(\Omega \tau, k_{ij} \tau)$, we see that varying the pulse area by changing $\tau$ or $\Omega$ leads, in general, to different results. Only in the limit of a full blockade, according to Eq. (\ref{full_block}), does $P_{\rm exc}$  again become a function of the pulse area only. The average time dependencies, for both types of pulses, slowly decrease after reaching saturation (the first maximum). This slow  decrease  can be explained by the reduction of the bandwidth $\Gamma$ for longer pulse durations, i.e. the blockading due to interactions becomes more effective. The slow increase with omega, after reaching saturation, is expected to happen due to  higher laser power.
This figure shows that the excitation probabilities for Gaussian pulses are systematically lower than those for square pulses. This is also expected  because of the following somewhat oversimplified argument. From the point of view of mean-field theory, the effect of interactions is a level shift, which can be expressed as some effective detuning of the excited level. Ignoring the time dependence of such level shifts, and in the first approximation, the probabilities are given by the Fourier transform of the pulse envelope. However, the Fourier transform of a Gaussian is also a Gaussian so that  the Fourier components of a Gaussian vanish much faster for large detunings than the corresponding Fourier components of a square pulse (which fall off as $\sim1/\Delta$).

In the previous figures we have assumed the interactions between $70p_{3/2}$ Rydberg atoms and, for square pulses, pulse durations of 10 ns at most. However, the product $k_{ij} \tau$ can still be significantly increased, either by going to higher principal quantum numbers $n$ (increasing $k_{ij}$ due to the $n^{11}$ scaling of $C_6$) or using longer pulses. In practice both ways should probably be used because for high $n$ the diatomic energy levels become very close to each other and the interactions at short internuclear distances have complicated forms due to avoided crossings between potential curves. However, increasing $\tau$ also means decreasing $\Gamma$ so that the excitation of atoms at shorter internuclear separations can be ignored and thus the van der Waals form of interaction can still be appropriate. Actually, the approximations in the superatom approach are justifiable if such excitation can be ignored. In fact, the largest $ \kappa_{pq}$ have no influence on superatom interactions $k_{ij}$ because for such $\kappa_{pq}$, ideally, both atoms $p$ and $q$ should belong to the same superatom. The simulation shown in Fig. \ref{fig_co5} is obtained for the product $k_{ij} \tau$ 15 times larger than the value used in Figs.  \ref{fig_co1}-\ref{fig_co2}. This would roughly correspond to the interactions between $n=90$ Rydberg atoms with the same bandwidth $\Gamma$.   Both curves are obtained after averaging  over many random spatial distributions of atoms. The oscillations are obviously more pronounced here than in Fig. \ref{fig_co1}-\ref{fig_co2}. To see how robust these oscillations are against density fluctuations, we varied the number of atoms in the sample according to a Poissonian statistics. The dashed curves is obtained after this additional averaging over density fluctuations.  

\begin{figure}[t]
   \centerline{\epsfxsize=2.9in\epsfclipon\epsfbox{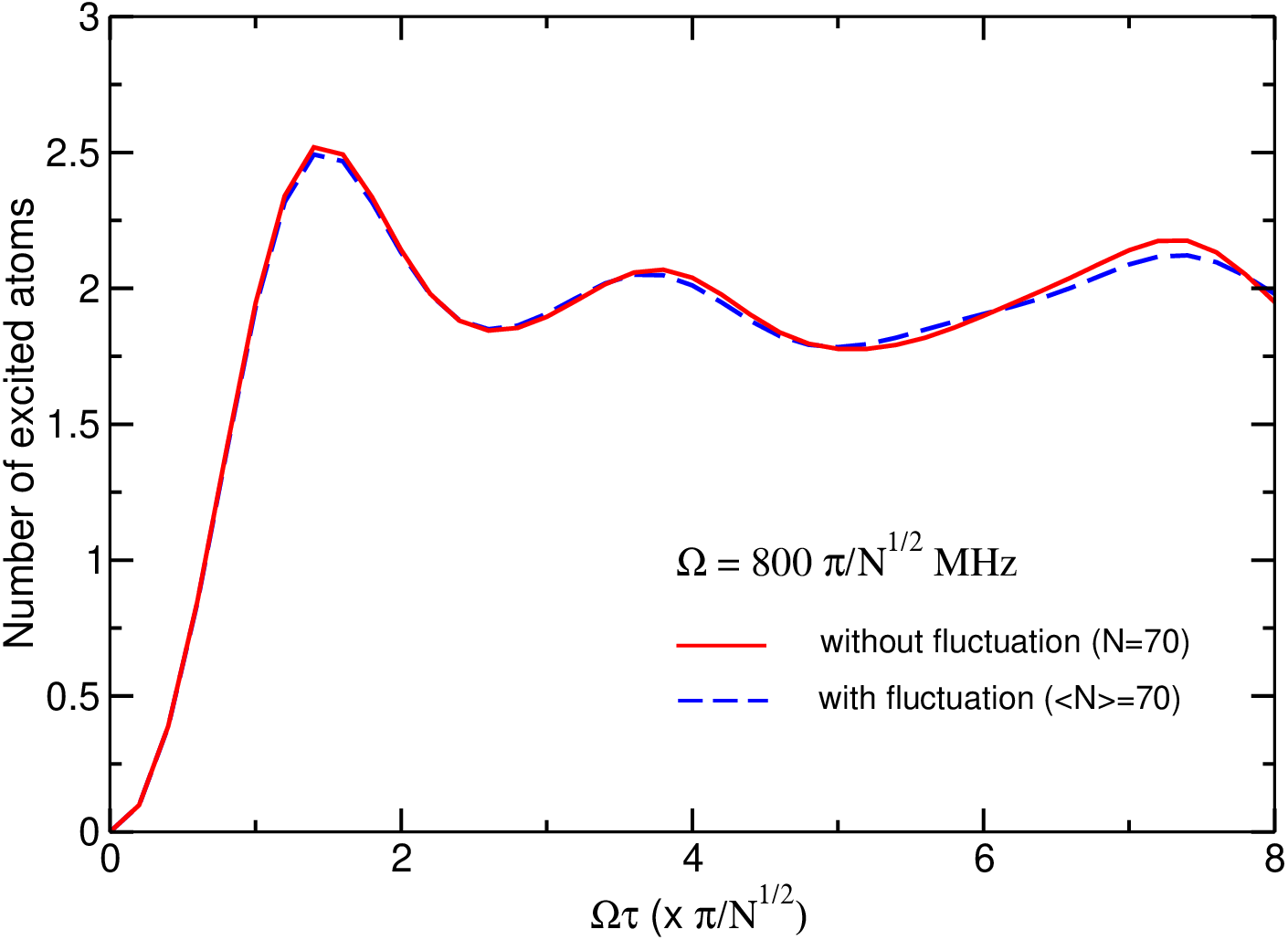}}
\caption{\label{fig_co5} The time dependence of the number of excited atoms for the scaled interaction strengths $k_{ij}$ 15 times larger than those used in the other simulations (Figs. \ref{fig_co1} and  \ref{fig_co2}). To check the robustness of the many-body oscillations against density fluctuations, the number of atoms in the sample is varied according to a Poissonian distribution. The solid curve is the dependence for $N= 70$ and the dashed curve is the average dependence over density fluctuations with $\langle N \rangle=70$.}
\end{figure}

In the first approximation, we can find an approximate parametrization of collective oscillations as follows. From the fact that the initial excitation probabilities do not depend on interactions, and the assumption that as soon as the interactions start to dominate the process they quickly saturate the excitations, one readily derives the following approximate formula
\begin{equation}\label{MBsaturation}
P_{\rm exc}^{\rm(1max)}=\frac{N_{\rm exc}^{\rm(1max)}}{N}=\frac{{F^{\rm(1max)}}^2}{4},
\end{equation}
where $P_{\rm exc}^{\rm(1max)}$ and $N_{\rm exc}^{\rm(1max)}$ are respectively the excitation probability and the number of excited atoms at the first maximum of $P_{\rm exc}$, and $F^{\rm(1max)}$ is the corresponding pulse area. The last formula is just the excitation probability for isolated (noninteracting) atoms if the pulse area is small. We can use it because, initially, there are no excited atoms and so there are no effects of interactions. In the next chapter we give a formal proof  (\ref{sigma_ee(2)}) for this claim. The effects of interactions for small pulse areas $F$ are proportional to $F^4$.  Depending on the strength of interactions, this saturation can be achieved much faster than the one in single-atom processes. In terms of frequency, the collective phenomena are faster than their counterparts  in single-atom processes. Assuming that the phase $\phi_{\rm coll}$ of collective oscillations can be simply characterized by some collective frequency $\Omega_{\rm coll}$ (i.e. $\phi_{\rm coll}\sim\Omega_{\rm coll}\tau$), where $\Omega_{\rm coll}=\alpha \Omega$, the scaling parameter $\alpha$ can be obtained from the saturation condition
\begin{equation}\label{alpha_sat}
\alpha F^{\rm(1max)}\approx\pi.
\end{equation}
Combining the last two equations one gets
\begin{equation}\label{alpha_solved}
\alpha\approx \sqrt{\frac{\pi}{4}}\sqrt{\frac{N_{\rm exc}^{\rm(1max)}}{N}}.
\end{equation}
To illustrate excitations in large samples with a significant excitation blockade, a domain  (or ``bubble'') picture is often used. Each domain represents a region in which there exists exactly one Rydberg atom. Denoting the number of atoms in a domain by $N_D$, then $N_D=N_{\rm exc}^{\rm(1max)}/N $, $\alpha=\sqrt{\pi/4}\sqrt{N_D}$ and  
\begin{equation}\label{alpha_Nd}
\Omega_{\rm coll}\approx \sqrt{\frac{\pi}{4}}\sqrt{N_D}\Omega.
\end{equation}

\begin{figure}
    \centerline{\epsfxsize=2.9in\epsfclipon\epsfbox{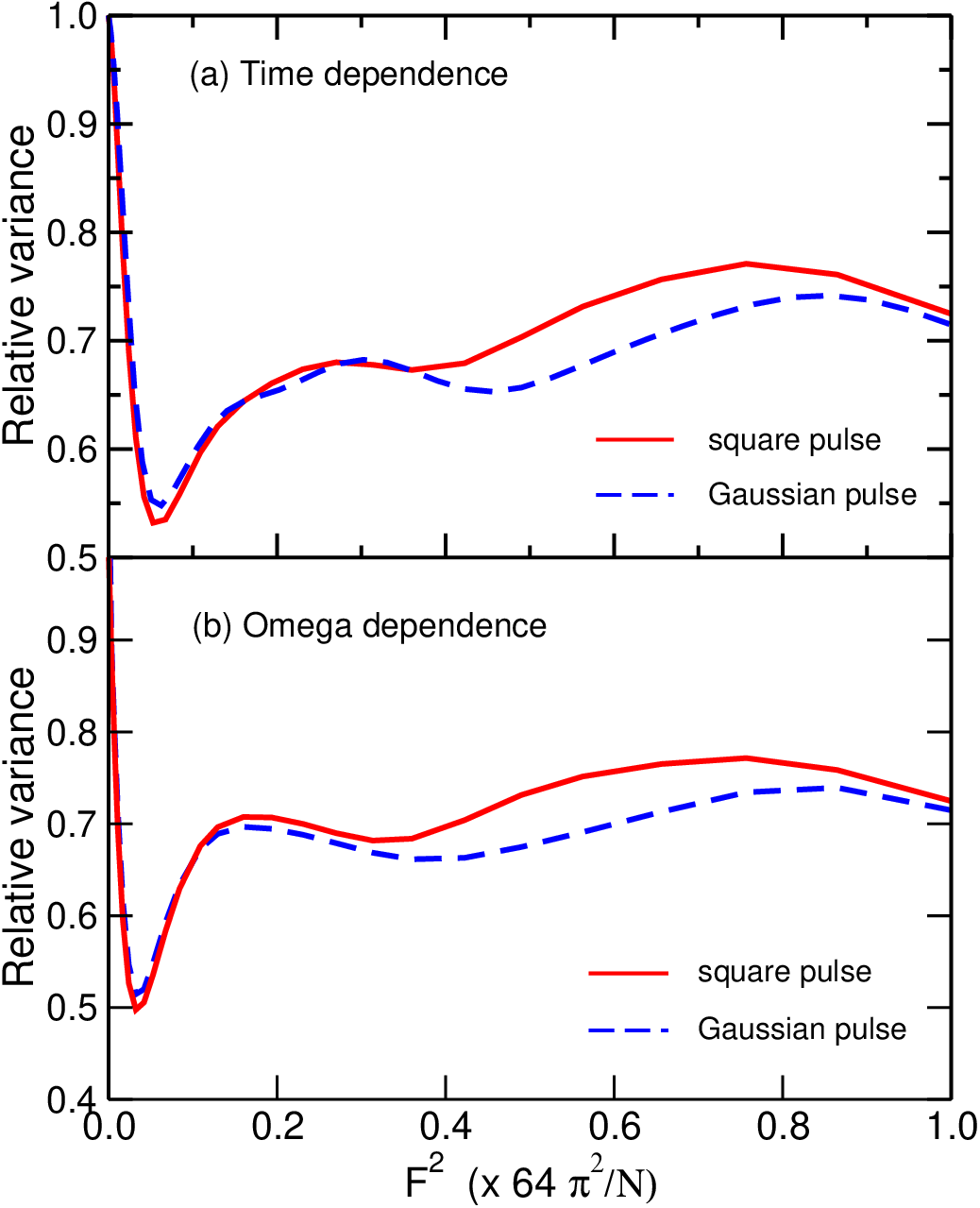}}
\caption{\label{fig_co3} The relative variance $\Delta n_{\rm exc}/\Delta n_{\rm exc}'$ of the number of excited Rydberg atoms versus the square of the pulse area for the same parameters used in Figs.  \ref{fig_co1}-\ref{fig_co2}. 
The actual variance $\Delta n_{\rm exc}$ is expressed using the calculated variance $\Delta n_{\rm exc}'$, obtained from $P$, assuming no correlation between excited atoms. In (a) pulse duration was varied and in (b) the single-atom Rabi frequency. Both solid lines correspond to square pulses and dashed lines to Gaussian ones.}
\end{figure}

The factor $\sqrt{\frac{\pi}{4}}=0.886...\approx 1$ can (and probably should) be approximated by one (because for $N_D=N$, according to Eq. (\ref{full_block}), $\Omega_{\rm coll}=\sqrt{N}\Omega$). We conclude that it seems that $\sqrt{N_D}$ is  the scaling factor of collective oscillations. We can verify this estimate by comparing with the numerical calculations shown in Fig. \ref{fig_co2} and Fig. \ref{fig_co5}. Also, if the phase of collective oscillations can be 
really described by a simple parameter $\Omega_{\rm coll}$ (or $\sqrt{N_D}$ ), then the ratio of the pulse areas corresponding to the second and the first maximum of $P_{\rm exc}$ should be three. The figures clearly show that the actual many-body behavior is much more complicated, so we just want to verify how much it deviates from the simple picture. For convenience, we define two parameters $\gamma=\alpha/\sqrt{N_D}$ and $\beta=F^{\rm(2max)}/F^{\rm(1max)}$. Interestingly, for all square pulse results presented in Figs. \ref{fig_co2} and \ref{fig_co5}, we find $\beta=2.5$, while for the Gaussian pulses in Fig. \ref{fig_co2}, $\beta=2.3$.  The parameter 
$\gamma$ varies more. For a square laser pulse in panels (a) and (b) in Fig. \ref{fig_co2}, $\gamma$ is respectively 0.95, 0.84 and 1.07. For the Gaussian ones in plots (a) and (b) in Fig. \ref{fig_co2}, $\gamma$ is 0.87 and 0.77.

Besides the many-body Rabi oscillations, there are other manifestations of correlations between interacting excited atoms. One can study the fluctuations of the number of excited atoms. Without these correlations, the probability to have a certain number of excited atoms is determined by the average excitation probability $P$  and the total number of atoms $N$. The probability $P(n_{\rm exc})$ to get any number $n_{\rm exc}\leq N$ of excited atoms is given by the Bernoulli formula
\begin{equation}\label{Bernouli}
P(n_{\rm exc})=\left(\begin{array}{c}
		N\\n_{\rm exc}
	   \end{array}\right)  P^{n_{\rm exc}}(1-P)^{N-n_{\rm exc}}.
\end{equation}
For a given excitation probability $P$, one can calculate  the expected variance $\Delta n_{\rm exc}'$ assuming no correlations, and then compare it with the actual $\Delta n_{\rm exc}$. We take the relative size of these variances as the measure of these fluctuations. This procedure can be done experimentally as well. In Fig. \ref{fig_co3}, we show the ratio $\Delta n_{\rm exc}/\Delta n_{\rm exc}'$ as a function of the pulse area. This ratio is smaller than one because there are some restrictions on which combinations of excited atoms can be likely excited in the presence of interactions. On the other hand, in the absence of correlations, all combinations of excited atoms are equally probable. This figure also shows that after reaching  the minimum, the ratio $\Delta n_{\rm exc}/\Delta n_{\rm exc}'$ increases again due to decoherence. The experimental results \cite{Liebisch} on counting statistics of somewhat larger samples than we consider here do show the sub-Poissonian character of Rydberg excitation.

\begin{figure}[t]
    \centerline{\epsfxsize=2.9in\epsfclipon\epsfbox{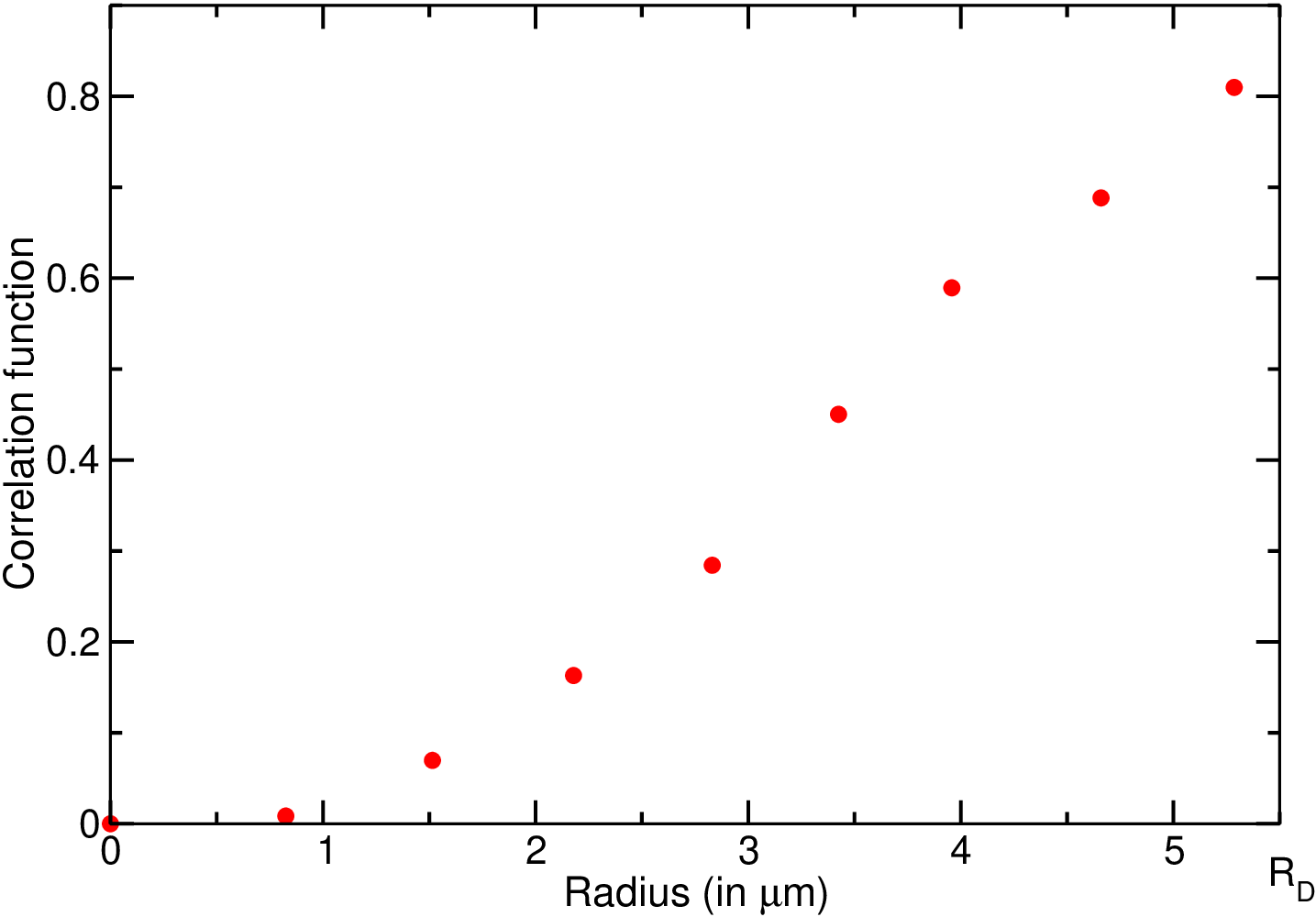}}
\caption{\label{fig_co4} The correlation function between a central atom and other atoms versus the distance between them for the same parameters used in Figs. \ref{fig_co1}-\ref{fig_co2}. This function is obtained by averaging over one hundred different random placements of 70 atoms. This means that 6900 atom pairs are used to get this dependence.}
\end{figure}

We have also calculated the spatial correlation function between the central atom and other atoms in  mesoscopic samples. For larger samples, this function was calculated in \cite{Hernandez}. They also found some interesting correlations if there is some frequency chirp of the laser pulse. In the presence of chirp, there is a region of internuclear separations $R$ for which the correlation function is greater than one. In the absence of chirp this region does not exist. We do not include any chirp in our simulations because it is known that it destroys even  the single-atom Rabi flopping. A delicate point for calculating the correlation function is that what we really calculate is the correlations beetwen superatoms and these superatoms are quite extended objects. Their linear size is about one third of the sample radius. Averaging over many random arrangements of atoms improves the determination of the correlation functions, but we do not insist on having  many points for this function (Fig. \ref{fig_co4}). In the supperatom approach, the correlation function $c(p,q)$ between any atom $p$ belonging to superatom $i$ and any atom $q$  belonging to superatom $j$ is the same for all $p$ and $q$ and equal to the correlation function $C(i,j)$ between superatoms $i$ and $j$. In addition, $c(p,q)=0$ for atoms $p$ and $q$ belonging to the same superatom. This suggests that the correlation function  $c(p,q)$ implicitly contains, in some sense, spatial averaging over the spatial extension of a superatom. Since, on average, excitation probabilities $P(i)$ depend on the atom's location, the correlation function has to be
\begin{equation}\label{cor_fun}
c(p,q)=\frac{P(p,q)}{P(p)P(q)},
\end{equation}
where $P(p,q)$ is the probability to excite simultaneously atoms $p$ and $q$.  In our case one of atoms $p$ and $q$ is always a central atom. The correlation function averaged over one hundred random positions of atoms is presented in Fig. \ref{fig_co4}. This figure shows that there are almost no correlations (i.e. $c(p,q)\approx1$) between the central atom  and the atoms near the sample surface.

\section{\label{sec:conclusionCO}Conclusion}
We have investigated the many-body dynamics of Rydberg excitation in ultracold mesoscopic systems of $\sim 10$ $\mu$m diameter. Various processes in  Rydberg gases are studied in detail by numerically solving the many-body Schr\"odinger equation. The possibility of many-body Rabi oscillation of Rydberg excitation is explored assuming that the $5s$ ground-state atoms are excited to the $70p$ Rydberg state by laser pulses of $\sim10$ ns duration. For typical experimental parameters, we have found that excitation probability for particular random distributions of atoms clearly exhibit many-body oscillations. However, when averaged over many distributions, these oscillations are significantly suppressed. More robust collective oscillations could be achieved by going to higher Rydberg states $n=90$, or decreasing the laser bandwidth (increasing pulse duration). We have evaluated the correlation function between a central atom and other atoms. It shows that excited atoms are strongly correlated  within a range of few $\mu$m.

\begin{acknowledgments}
 This research was funded by the National Science Foundation.
\end{acknowledgments}

\end{document}